\documentclass[12pt]{article} 
\usepackage{epsfig} 
\usepackage{amsfonts} 
\usepackage{amssymb}
\usepackage{amsmath}  
\usepackage{amsbsy}  
\usepackage{wasysym}
\usepackage{cite}
\usepackage{multirow} 
\input axodraw.sty
\catcode`\@=11 
\@addtoreset{equation}{section} 
\def\theequation{\thesection.\arabic{equation}} 
\@twosidetrue 
\catcode`\@=11  
\def\section{\@startsection{section}{1}{\z@}{3.5ex plus 1ex minus 
.2ex}{2.3ex plus .2ex}{\large\bf}} 
\def\thesection{\arabic{section}} 
\def\thesubsection{\arabic{section}.\arabic{subsection}} 
\def\thesubsubsection{\arabic{section}.\arabic{subsection}.\arabic{subsubsection}} 
\def\appendix{\setcounter{section}{0} 
\def\thesection{\Alph{section}} 
\def\theequation{\Alph{section}.\arabic{equation}} 
\def\thesubsection{\Alph{section}.\arabic{subsection}} 
\def\thesubsubsection{\Alph{section}.\arabic{subsection}.\arabic{subsubsection}} 
\def\section{\@startsection{section}{1}{\z@}{3.5ex plus 1ex minus 
   .2ex}{2.3ex plus .2ex}{\large\bf}} } 
\newcount\minutes 
\newcount\scratch 
\def\timestamp{%
\scratch=\time 
\divide\scratch by 60 
\edef\hours{\the\scratch} 
\multiply\scratch by 60 
\minutes=\time 
\advance\minutes by -\scratch 
---$\,$\hours:\null 
\ifnum\minutes< 10 0\fi 
\the\minutes} 

\def\sla#1{\ifmmode%
\setbox0=\hbox{$#1$}%
\setbox1=\hbox to\wd0{\hss$/$\hss}\else%
\setbox0=\hbox{#1}%
\setbox1=\hbox to\wd0{\hss/\hss}\fi%
#1\hskip-\wd0\box1 } 
\def\wwa{{$W^+W^-\gamma$}}
\def\zza{{$ZZ\gamma$}}
\def\vbfnlo{{\tt VBFNLO}}

\def\beq{\begin{equation}} 
\def\eeq{\end{equation}} 
 
\def\beqn{\begin{eqnarray}} 
\def\eeqn{\end{eqnarray}}

\def\({\left(} 
\def\){\right)} 
\def\as{\ifmmode \alpha_s \else $\alpha_s$ \fi}

\begin{document} 
\begin{titlepage} 
\nopagebreak 
{\flushright{ 
        \begin{minipage}{5cm}
         KA--TP--11--2009  \\            
         SFB/CPP-09-105\\ 
         FTUV--09--1101 \\
         IFUM-938-FT \\
        \end{minipage} } 

} 
\vfill 
\begin{center} 
{\LARGE \bf 
 \baselineskip 0.5cm 
NLO QCD corrections to $W^+W^-\gamma$ and $ZZ\gamma$ production with leptonic decays} 
\vskip 0.5cm  
{\large   
G. Bozzi$^1$, F. Campanario$^{2,3}$, V. Hankele$^{2}$ and D. Zeppenfeld$^2$
}   
\vskip .2cm
{$^1$ {\it Dipartimento di Fisica, Universit\`a di Milano and INFN,
    Sezione di Milano\\ Via Celoria 16, I-20133 Milano, Italy} 
    }\\
{$^2$ {\it Institut f\"ur Theoretische Physik, Karlsruhe Institute of Technology,~~~~~~~~~~~ \\  Universit\"at Karlsruhe, 76128 Karlsruhe, Germany}
    }\\
{$^3$ {\it Departament de F\'isica Te\`orica and IFIC, Universitat de 
    Val\`encia - CSIC,~~\\ E-46100, Burjassot, Val\`encia, Spain}  }\\
 \vskip 1.3cm     
\end{center} 
\nopagebreak 
\begin{abstract}
The computation of the ${\cal O}(\alpha_s)$ QCD corrections to the cross
sections for $W^+W^-\gamma$ and $ZZ\gamma$ production in hadronic
collisions is presented. We consider the case of a real photon in the
final state, but include full leptonic decays of the $W$ and $Z$ bosons.
Numerical results for the LHC and the Tevatron are obtained
through a fully flexible parton level Monte Carlo based on the structure of
the VBFNLO program, allowing an easy implementation of arbitrary cuts
and distributions. 
We show the dependence on scale variations of the integrated cross
sections and provide evidence that NLO QCD
corrections strongly modify the LO predictions for observables at the
LHC both in magnitude and in shape.
\end{abstract} 
\vfill 
\hfill 
\vfill 

\end{titlepage} 
\newpage               
%
%
\section{Introduction}
\label{sec:intro}
The experimental precision that will be reached in cross section 
measurements at the CERN Large Hadron Collider (LHC) demands an effort of the
theoretical community in providing accurate phenomenological predictions.
NLO QCD corrections for cross sections and distributions have thus
become mandatory, and many 
relevant processes at hadron colliders are now known to this accuracy.

Events with multiple gauge bosons in the final state provide an
irreducible background to many new physics searches (see, for instance,
\cite{Campbell:2006wx} for a discussion of the relevant backgrounds in
the search of New Physics at the LHC).  In addition, the
triple gauge couplings involved in the contributing diagrams allow for
restrictive tests of the gauge sector of the Standard Model. The
process $pp \to$ \wwa~+~X is particularly important since it is also
sensitive to four gauge boson couplings, namely the $WWZ\gamma$ and
$WW\gamma\gamma$ vertices~\cite{quartic}.

In this paper, we present the ${\cal O}(\alpha_s)$ corrections for the
processes 
\begin{eqnarray}
\label{processes}
"W^+W^-\gamma" \qquad & pp,\; p\bar p \to 
\nu_{l_1} l_1^+ \bar\nu_{l_2} l_2^- \gamma~+X  \nonumber \\ 
"ZZ\gamma" \qquad & pp,\; p\bar p \to l_1^+ l_1^- l_2^+ l_2^- \gamma~+X. 
\end{eqnarray}
Similar to previous work on triple weak boson production
\cite{Lazopoulos:2007ix,Hankele:2007sb,Campanario:2008yg,Binoth:2008kt},
we find that the QCD corrections are sizeable and also modify the shape
of the differential distributions for many observables: this proves that
a simple rescaling of the LO results is not adequate and a full NLO Monte
Carlo is needed for a quantitative determination of quartic
couplings at the LHC. We have implemented our calculation within the
VBFNLO framework \cite{Arnold:2008rz}, a fully-flexible parton level 
Monte Carlo
program which allows the definition of arbitrary acceptance cuts and distributions.

The paper is organized as follows: in Section \ref{sec:calc} we provide
an example of the relevant Feynman diagrams at tree level, a short
account of the strategies used to compute the real and virtual
corrections and the various checks performed both internally and against
other available codes. In Section \ref{sec:res} we show numerical
results, including the scale variations of the LO and NLO integrated
cross sections and some selected differential distributions. Conclusions
are given in Section \ref{sec:concl}. 
%
%
\section{The calculation}
\label{sec:calc}
\begin{figure}[hbtp]
\begin{center}
\setlength{\unitlength}{1pt}
\begin{picture}(50,110)(0,0)

\ArrowLine(-195,15)(-160,50)
\ArrowLine(-195,85)(-160,50)
\Photon(-160,50)(-110,50){3}{5}
\Photon(-110,50)(-60,85){3}{5}
\Photon(-110,50)(-60,50){3}{5}
\Photon(-110,50)(-60,15){3}{5}
\ArrowLine(-50,95)(-60,85)
\ArrowLine(-60,85)(-50,75)
\ArrowLine(-50,25)(-60,15)
\ArrowLine(-60,15)(-50,5)
\put(-145,60){$Z,\gamma$}
\put(-95,75){$W$}
\put(-70,60){$\gamma$}
\put(-95,10){$W$}
\put(-140,-5){a)}

\ArrowLine(-20,15)(30,15)
\ArrowLine(30,85)(-20,85)
\ArrowLine(30,15)(30,85)
\Photon(30,15)(90,15){3}{5}
\Photon(30,85)(60,85){3}{3}
\Photon(60,85)(90,105){3}{3}
\Photon(60,85)(90,65){3}{3}
\ArrowLine(90,105)(100,115)
\ArrowLine(100,95)(90,105)
\ArrowLine(90,65)(100,55)
\ArrowLine(100,75)(90,65)
\put(35,95){$Z,\gamma$}
\put(70,105){$W$}
\put(70,55){$W$}
\put(70,25){$\gamma$}
\put(30,-5){b)}

\ArrowLine(135,15)(185,15)
\ArrowLine(185,85)(135,85)
\ArrowLine(185,15)(185,50)
\ArrowLine(185,50)(185,85)
\Photon(185,85)(235,85){3}{5}
\Photon(185,50)(235,50){3}{5}
\Photon(185,15)(240,15){3}{5}
\ArrowLine(245,95)(235,85)
\ArrowLine(235,85)(245,75)
\ArrowLine(245,60)(235,50)
\ArrowLine(235,50)(245,40)
\put(210,95){$W$}
\put(210,60){$W$}
\put(230,25){$\gamma$}
\put(185,-5){c)}
\end{picture}  \\
\setlength{\unitlength}{1pt}
\caption{Examples of the three topologies of Feynman diagrams
  contributing to the process $pp\to$\wwa + X \, at tree-level.}
\label{fig:1}
\end{center}
\end{figure}
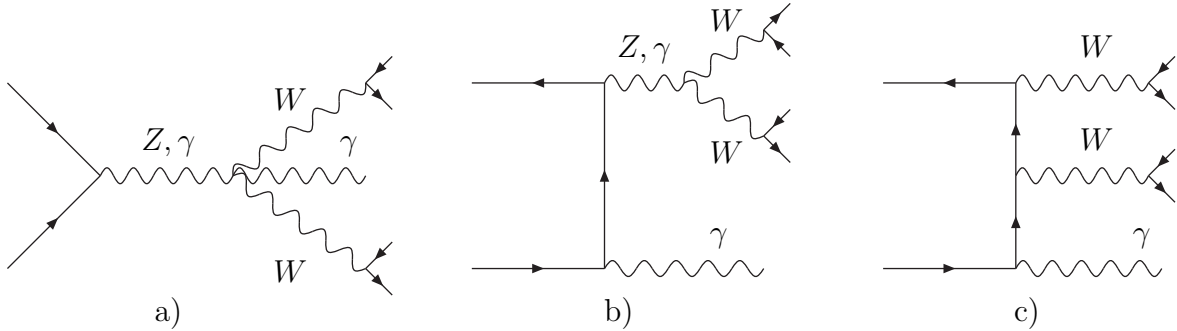

We consider the processes (\ref{processes}) up to order $\as\alpha^6$ in
the limit where all fermions are massless. Among the LO diagrams (110 in
the ``\wwa'' case, 336 in the ``\zza'' case) we can distinguish three
different topologies: they correspond to the cases when 1, 2, or 3 vector bosons are attached to the quark line
(respectively case a), b), or c) of Fig.\ref{fig:1}, where examples for
``\wwa'' production are shown. The single and double vertex topologies
of Fig.\ref{fig:1} a) and b) also exist for ``\zza'' production since we
include off-shell effects in our calculation, i.e. the photon or a 
$Z$-boson can be radiated off a final state lepton.) 

In all cases, the leptonic decays of the $W$ and $Z$ bosons and
combinations of subgraphs corresponding to decays of the same virtual
$W$, $Z$, or photon are
factorized in the form of {\it leptonic tensors} and computed
independently from the rest of the cross section, in analogy with the
procedure in \cite{Jager:2006zc}: this
reduces computational time since the same decay current
may appear in many Feynman diagrams and in different subprocesses and
can, thus, be computed once for each
phase-space point and saved for later use.
For the computation of the matrix elements, we use the helicity method
introduced in \cite{Hagiwara:1985yu}.

The cross section at order \as contains contributions from real
emission and  virtual corrections from the interference of one-loop
diagrams with the Born amplitude. We use the Catani-Seymour dipole
subtraction method \cite{Catani:1996vz} to handle the cancellation of
infrared divergences between real emission and virtual
corrections. Factorization  of initial state collinear singularities
into the parton distribution functions leads to additional 
``finite collinear terms''.

The NLO real corrections are generated by the emission of a gluon off
the quark line or by the emission of a quark through a $g \to q\bar q$
splitting in the initial state. The strategy of computing {\it leptonic
  tensors} separately proves to be particularly useful in this case,
since we have a total of 272 (896) different Feynman diagrams
contributing to the ``\wwa'' (``\zza'') process when attaching an
additional gluon to the Born diagrams. 

In the case of real emissions, the definition of an isolated on-shell
photon requires further discussion. Photon emission collinear to a
massless quark results in an additional singularity, which needs to be
regularized.  The simple rejection of events
containing partons inside a cone of fixed width drawn around the photon
is not a permissible choice since it will suppress the phase space for
soft gluon emissions and spoil the cancellation of infrared
divergences. In this paper we avoid the complications of quark
fragmentation to photons by using an effective method which has been
proposed by Frixione \cite{Frixione:1998jh}. The precise definition of
this {\it photon isolation cut} is discussed in Section \ref{sec:res}.

The NLO virtual corrections come from the interference of leading order
diagrams with one-loop diagrams obtained by attaching a gluon to the
quark-antiquark line. In the case where only one vector boson is
attached to the quark line, i.e. topology  a), one only has vertex
corrections, while diagrams belonging to topologies  b) and  c)
give rise to boxes and pentagons. 
The tensor coefficients of the loop integrals have been computed  \`a la
Passarino-Veltman~\cite{Passarino:1978jh} up to box level, while the
Denner-Dittmaier reduction 
procedure \cite{Denner:2002ii} has been used for the pentagons, 
which avoids the problem of small Gram determinants occurring
in pentagons with planar configurations of the external
momenta. Denner-Dittmaier reduction is
now fully implemented in the public version of the \vbfnlo \, code.
The overall contribution from virtual corrections can be written as
\begin{align}
M_V = \widetilde{M}_V + \ \frac{\alpha_S}{4 \pi} \ C_F \ \left( \frac{4
      \pi \mu^2}{Q^2} \right)^\epsilon  \ \Gamma{(1 + \epsilon)} \ \left[
    -\frac{2}{\epsilon^2} - \frac{3}{\epsilon} - 8 + \frac{4 \ \pi^2}{3}
  \right] \ M_B, \label{eq:MV}
\end{align}
where $M_B$ is the Born amplitude and $Q$ is the partonic center-of-mass
energy, i.e. the invariant mass of the final state $VV\gamma$
system. The term $\widetilde{M}_V$ is the finite part of the virtual 
corrections to 2 and 3 weak boson amplitudes as in Fig.\ref{fig:1}b) and c)
which we call
boxline and pentline contributions in the following, so named since boxes and 
pentagons constitute the most complex loop diagrams, respectively.

We have verified analytically and numerically that in order to obtain the
above factorization formula of the infrared divergences against the born
amplitude, the transversality property of the photon ($\epsilon(p_\gamma)\cdot
p_\gamma= 0$) must be used, otherwise, additional IR singularities which are 
not proportional to
the born amplitude do appear. After checking that there are no additional
$1/\epsilon$ terms due to the massless on-shell photon, the analytical
structure of $\widetilde{M}_V$ can be obtained from the massive weak boson 
case considered in \cite{Campanario:2008yg}. Nevertheless, we have 
recomputed equivalent pentline and boxline contributions for massless 
particles and found that they agree at the double precision accuracy 
level with those of~\cite{Campanario:2008yg}, once the proper scalar integrals
are used.


A further reduction of computing time is obtained for ``\wwa'' production
by implementing a trick already used in
\cite{Jager:2006zc,Hankele:2007sb,Campanario:2008yg}.
We write the polarization vector of the vector bosons as
\begin{align}
\label{mapping}
\epsilon_{V}^\mu  = x_{V} \, q_{V}^\mu + \tilde{\epsilon}_{V}^\mu
\end{align}
where $q_V$ is the momentum of the massive boson ($V=W^\pm $). A pentagon
contracted with one of the external momenta $q_V$ can always be reduced
to a difference of box diagrams. Thus we effectively isolate
a {\it true} pentagon contribution and shift the rest of the 5-point
contribution to the less time-consuming boxes. The remainder 
$\tilde{\epsilon}_{V}$ is chosen in such a way that 
\begin{align}
\tilde{\epsilon}_{V} \cdot (q_{V_1} + q_{V_2}) = 0 \, ,
\end{align}
i.e., the time-component of the shifted polarization vector is zero in
the center-of-mass system of the V-pair. This choice yields particularly 
small pentline contributions which, hence, can be determined with lower 
Monte Carlo statistics. For ``\zza'' production, the pentline contributions 
group into gauge invariant subsets which are invariant under the replacement 
of Eq.(\ref{mapping}). Thus the trick does not speed up the code but 
provides a consistency check for the calculation.

We have performed a number of checks on our final results. First, we
compared all tree level matrix elements (including the ones for real emission
corrections) against amplitudes generated by 
{\tt MadGraph}\cite{Stelzer:1994ta}: the agreement is
at the level of machine precision ($10^{-15}$). Second, we have compared
the LO cross sections both for $pp \to $ \wwa$~+X$ and $pp \to $ \zza$~+X$
against {\tt HELAC}\cite{Cafarella:2007pc} and found agreement at the
per mill level. Third, we have compared integrated tree level cross sections for
$pp(p\bar p)
\to $ \wwa$(j)~+X$ and $pp(p \bar p) \to $ \zza$(j)~+X$ against {\tt MadEvent} and
{\tt Sherpa}~\cite{Sherpa:09} finding agreetment at the per mill level.
Finally, 
we also performed
gauge tests based on Ward identities for the NLO virtual corrections at
different levels of complexity.
%
%

\section{Results}
\label{sec:res}

Numerical results have been obtained through the implementation of
our calculation into a NLO Monte Carlo program based on the structure of
the \vbfnlo \, code \cite{Arnold:2008rz}. This and other processes with
a real photon in the final state will be included in the public version
of \vbfnlo \, in a future release. 

The value of the total cross section for ``\zza'' production, including
leptonic Z-decays, turns out to
be less than a femtobarn at LHC energies: since it will probably be too
hard to detect, at least in an early luminosity run, in this section we
focus on the $pp\to$ \wwa$+X$ process. 

For the electroweak parameters we use the $W$ and $Z$ boson masses and 
the Fermi constant as input. From these we derive the electromagnetic 
coupling and the weak mixing angle via tree level relations, i.e. we use 
\begin{align}
&m_W = 80.398 \ \mathrm{GeV}  &&\sin^2{(\theta_W)} = 0.22264 \nonumber\\
&G_F = 1.16637 \cdot 10^{-5} \ \mathrm{GeV}^{-2} 
&&\alpha^{-1} = 132.3407 \; .  \label{eq:ew}
\end{align}
We do not consider bottom and top quark effects. The remaining quarks are
assumed to be massless and we work in the approximation where 
the CKM matrix is the identity matrix.
We choose the invariant ``\wwa'' mass as the central value for the
factorization and renormalization scales: 
\begin{align} \label{eq:WWAmass}
\mu_F = \mu_R = \mu_0 = \sqrt{(p_{\ell_1} + p_{\ell_2} + p_{\nu_1} +
  p_{\nu_2} + p_{\gamma})^2}.
\end{align}
We use the CTEQ6L1 parton distribution function at LO and the CTEQ6M
set with $\alpha_S(m_Z)=0.1176$ at NLO \cite{Pumplin:2002vw}.

Since the photon can be emitted either from the initial quark line or
from each of the two final state charged leptons, we divide the phase
space into three separate regions to consider all the possibilities and
then sum the contributions to get the total result. The regions are 
generated as (approximately) on-shell $W^+\to e^+\nu_e\gamma$ and 
$W^-\to \mu^-\bar\nu_\mu\gamma$ three-body decay and triple electroweak 
boson production, respectively. 
We impose a set of minimal cuts on the rapidity, $y_{\gamma(\ell)}$, and
the transverse momenta , $p_{T_{\gamma(\ell) }}$, of the photon and
the charged leptons which are designed to represent typical experimental 
requirements. Furthermore, leptons, photons and jets must be well separated
in the rapidity-azimuthal angle plane. 
Specifically, the cuts we impose are
\begin{equation}
p_{T_{\gamma(\ell)}} > 20 \ \mathrm{GeV} \qquad
|y_{\gamma(\ell)}| < 2.5 \qquad
R_{\ell\gamma} > 0.4  \qquad
R_{j \ell} > 0.4  \qquad
R_{ j \gamma} > 0.7 
\label{eq:cuts}
\end{equation}
where, in our simulations, a jet is defined as a colored parton of 
transverse momentum $p_{Tj}>20$ GeV.
For photon isolation, we implement the procedure 
introduced in \cite{Frixione:1998jh}: if $i$ is a parton with transverse energy
$E_{T_i}$ and a separation $R_{i\gamma}$ with a photon of transverse momentum
$p_{T_\gamma}$, then the event is accepted only if
\beq
\Sigma_i \, E_{T_i} \, \theta (\delta - R_{i\gamma}) \, \leq \, p_{T_\gamma} \,
\frac{1-\cos\delta}{1-\cos\delta_0} \,\,\,\,\,\,\,\,\,\,
(\mathrm{for\,all} \,\,\,\,\,\delta\leq\delta_0) \label{eq:isol}
\eeq
where $\delta_0$ is a fixed separation that we set equal to 0.7. A quick look
at Eq.(\ref{eq:isol}) reveals that a sufficiently soft gluon can be 
arbitrarily close
to the photon axis, while the energy of an exactly collinear parton must be
vanishing in order to pass the isolation cut. Collinear-only events
(leading to fragmentation contributions) are thus rejected, while soft
emissions are retained, as desired. Additionally, for ``\zza'' production 
we impose
that the invariant mass of any combination of two charged leptons,
$m_{\ell \ell}$, be larger than 15 GeV in order to avoid
virtual-photon singularities in $\gamma^*\to l^+ l^- $ at a low $m_{\ell
\ell}$.
In Table~\ref{LHCnumbers}, we give results for the integrated cross 
sections for ``\wwa'' and ``\zza'' production at the LHC for the 
given cuts as well as for a harder cut on the photon transverse momentum
of $p_{T_\gamma}>30 $ GeV. In
Table~\ref{Tevnumbers}, the numbers for the Tevatron are presented for
the cuts as in Eqs.(\ref{eq:cuts},\ref{eq:isol}), but for less restrictive 
transverse momentum cuts, $p_{T_{\gamma(\ell)}}> 10(10)$ GeV and 
$p_{T_{\gamma(\ell)}}> 20(10)$ GeV. The center-of-mass energy is set
to $14$ TeV for the LHC and $1.96$ TeV for Tevatron collisions, respectively.

\begin{table}[ht]
\begin{center}
\renewcommand{\arraystretch}{1.15}
\begin{tabular*}{0.85\textwidth}{@{\extracolsep{\fill}}|c@{\hspace{1cm}}|c@{\hspace{1.2cm}}|c@{\hspace{1.2cm}}|}
\hline
LHC &  LO  [fb]     &   NLO [fb] \\ \hline
$~~\sigma("WW\gamma" \to e^+ \nu \mu^- \bar \nu \gamma)$
&\multirow{2}{*}{1.695} &\multirow{2}{*}{2.881  
}\\ 
$~~p_{T_{\gamma(\ell)}}>20(20)$ GeV 
& &\\ \hline

$~~\sigma("WW\gamma" \to e^+ \nu \mu^- \bar \nu \gamma)$
&\multirow{2}{*}{0.9580 } &\multirow{2}{*}{1.738 }\\ 
$~~p_{T_{\gamma(\ell)}}>30(20)$ GeV 
& &\\ \hline

$~~\sigma("ZZ\gamma" \to e^+ e^- \mu^+ \mu^-\gamma)$
&\multirow{2}{*}{0.07786 } &\multirow{2}{*}{0.1062}\\ 
$~~p_{{T_{\gamma(\ell)}}}>20(20)$ GeV  
& &\\ \hline

$~~\sigma("ZZ\gamma" \to e^+ e^- \mu^+ \mu^-\gamma)$
&\multirow{2}{*}{0.03969} &\multirow{2}{*}{0.05577 }\\ 
$~~p_{{T_\gamma(\ell)}}>30(20)$ GeV 
& &\\ \hline
 \end{tabular*}
\caption[]{Total cross sections at the LHC for $pp \to $ \wwa$+X$ and
  $pp\to $\zza$+X$ with leptonic decays, at LO and NLO, and for two 
  sets of cuts. Relative
  statistical errors of the Monte Carlo are below $10^{-3}$.}
\label{LHCnumbers}
 \end{center}
 \end{table}

\begin{table}[ht]
\begin{center}
\renewcommand{\arraystretch}{1.15}
\begin{tabular*}{0.85\textwidth}{@{\extracolsep{\fill}}|c@{\hspace{1cm}}|c@{\hspace{1.2cm}}|c@{\hspace{1.2cm}}|}

\hline
Tevatron &  LO [fb]       &   NLO [fb] \\ \hline
$~~\sigma("WW\gamma" \to e^+ \nu \mu^- \bar \nu \gamma)$
&\multirow{2}{*}{0.9015 } &\multirow{2}{*}{1.271  }\\ 
$~~p_{T_{\gamma(\ell)}}>10(10)$ GeV & &\\ \hline

$~~\sigma("WW\gamma" \to e^+ \nu \mu^- \bar \nu \gamma)$
&\multirow{2}{*}{0.3755 } &\multirow{2}{*}{0.5342 }\\ 
$~~p_{T_{\gamma(\ell)}}>20(10)$ GeV  & &\\ \hline

$~~\sigma("ZZ\gamma" \to e^+ e^- \mu^+ \mu^-\gamma)$
&\multirow{2}{*}{0.05469} &\multirow{2}{*}{0.07681}\\ 
$~~p_{T_{\gamma(\ell)}}>10(10)$ GeV & &\\ \hline

$~~\sigma("ZZ\gamma" \to e^+ e^- \mu^+ \mu^-\gamma)$
&\multirow{2}{*}{0.02440 } &\multirow{2}{*}{0.03441}\\ 
$~~p_{T_{\gamma(\ell)}}>20(10)$ GeV & &\\ \hline
 \end{tabular*}
\caption[]{Total cross sections at the Tevatron for $p\bar p \to $ \wwa$+X$ and
  $p\bar p\to $\zza$+X$ with leptonic decays, at LO and NLO, and for two 
  sets of cuts. Relative
  statistical errors of the Monte Carlo are below $10^{-3}$.}
\label{Tevnumbers}
 \end{center}
 \end{table}
In the following we show \wwa\, results for the process $pp \to \nu_{e}
e^+ \bar\nu_{\mu} \mu^- \gamma+X$, i.e we do not consider interference 
effects due to identical leptons, which generally will be small. The
cross sections have been multiplied by a combinatorial factor of 4, 
however, and correspond to the production of any combination of 
electrons or muons.

\begin{figure}[tbp]
\includegraphics[scale = 1]{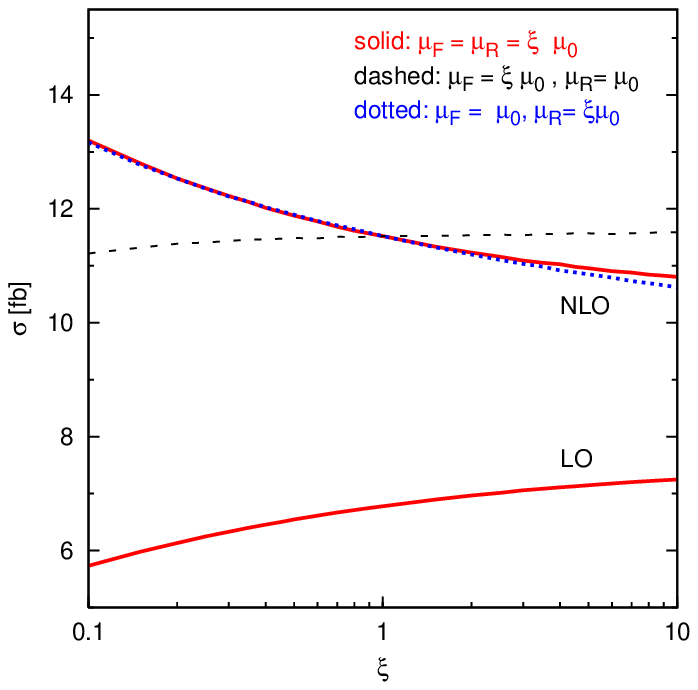}
\includegraphics[scale = 1]{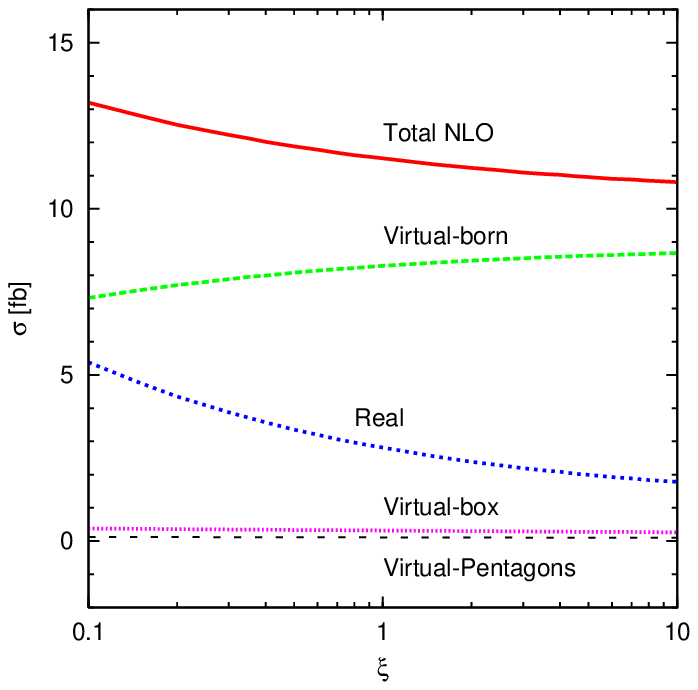}
\caption[]{\label{fig:2}
{\it Left:} {\sl Scale dependence of the total LHC cross section for 
  $p p \to W^+W^-\gamma+X \to \ell^+ \ell^- \gamma +\sla{p}_T+X$ at LO and
  NLO within the cuts of Eqs.~(\ref{eq:cuts},\ref{eq:isol}).
  The factorization and renormalization scales are together or
  independently varied in the range from $0.1 \cdot \mu_0$ to $10 \cdot
  \mu_0$.} 
{\it Right:} {\sl Same as in the left panel but for the different NLO
  contributions at $\mu_F=\mu_R=\xi\mu_0$.}}
\end{figure}

In order to
estimate the scale uncertainty of the total cross section, we study the
variation of the numerical results in the interval 
\beq
\mu_F, \mu_R = \xi\cdot \mu_0 \,\,\,\,\,\,\,\,\,\,\,\,\, (0.1 < \xi < 10),
\label{eq:scale}
\eeq
where $\mu_0$ is the ``\wwa'' or ``\zza'' invariant mass. The variation
of the total ``\wwa'' cross section at LO and NLO is shown on 
the left panel of Figure \ref{fig:2}. 
It is evident that the LO scale variation 
greatly underestimates the
size of NLO corrections, which give a K-factor of 1.70 for
$\mu_F=\mu_R=\mu_0$. The NLO scale uncertainty is about 6$\%$ when varying
the factorization and the renormalization scale $\mu=\mu_F=\mu_R$ up
and down by a factor 2 around the reference scale $\mu_0=Q$, slightly
smaller than for the LO uncertainty, 
and is dominated by the dependence on $\mu_R$,
which gives a negative slope with increasing energy, while it has a
quite flat dependence on $\mu_F$.

On the right panel of Figure \ref{fig:2}, the scale dependence and relative
size of the different NLO contributions are shown. One clearly sees that
almost the entire scale variation of the integrated NLO cross section is accounted
for by the real emission contributions, defined here as the real emission
cross section minus the Catani-Seymour subtraction terms plus the finite 
collinear terms. The virtual contributions proportional to the Born matrix element 
(due to the $M_B$ term in Eq.~(\ref{eq:MV}))
constitute the bulk of the NLO corrections, being roughly
twice as large as the real emissions, while the finite virtual remainders
due to boxes and pentagons only represent 3\% and less than 1\% of the
total result, respectively, and their scale dependence is basically flat.

\begin{figure}[tbp]
\includegraphics[scale = 1]{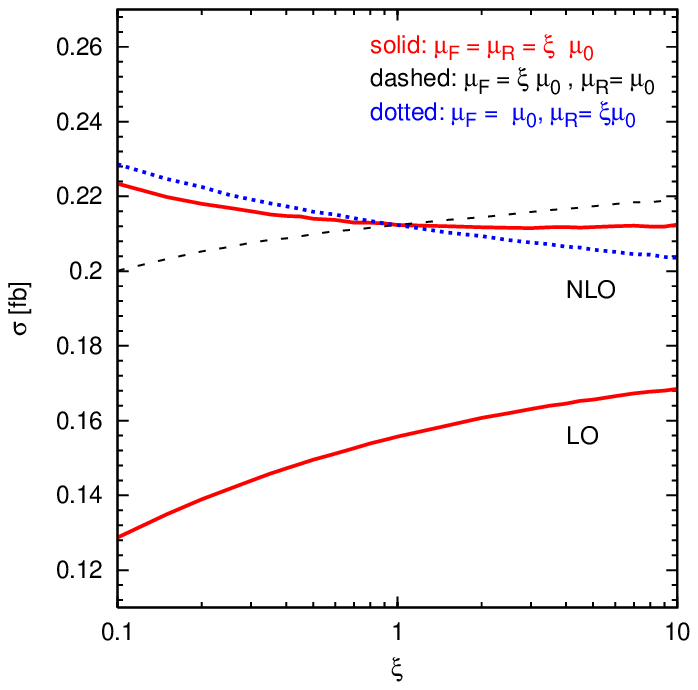}
\includegraphics[scale = 1]{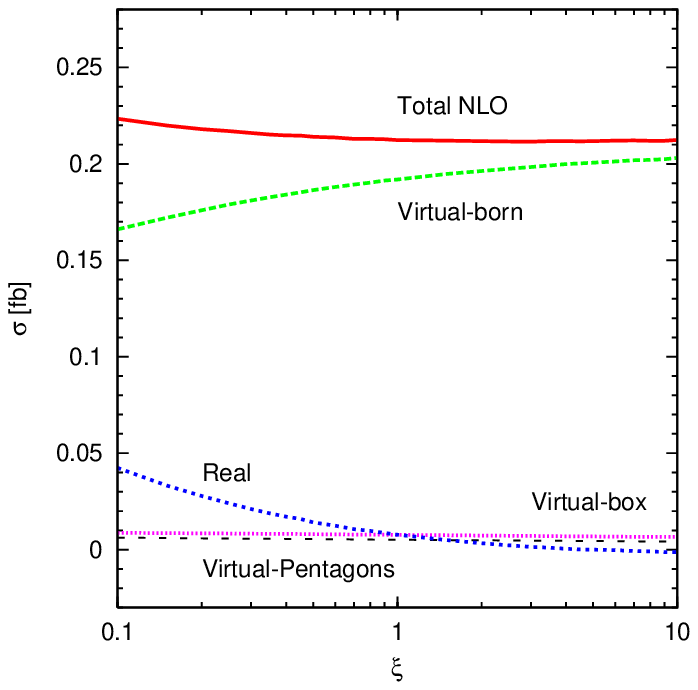}
\caption[]{\label{fig:21}
{\it Left:} {\sl Scale dependence of the total LHC cross section for 
  $p p \to ZZ\gamma+X \to \ell^+_1 \ell^-_1\ell^+_2 \ell^-_2  \gamma+X $ 
  at LO and NLO within the cuts of Eqs.~(\ref{eq:cuts},\ref{eq:isol}).
  The factorization and renormalization scales are together or
  independently varied in the range from $0.1 \cdot \mu_0$ to $10 \cdot
  \mu_0$.} 
{\it Right:} {\sl Same as in the left panel but for the different NLO
  contributions at $\mu_F=\mu_R=\xi\mu_0$.}}
\end{figure}

Similar considerations apply for ``\zza'' production where a
combinatorial factor of 2 has been included for $p p \to ZZ\gamma+X \to \ell^+_1 \ell^-_1\ell^+_2 \ell^-_2  \gamma+X$. In Figure \ref{fig:21}, one
observes that the overall scale variation decreases when
varying
the factorization and the renormalization scale $\mu=\mu_R=\mu_0$ up
and down by a factor 2 around the reference scale $\mu_0=Q$ from roughly
$\pm$ 7\% at LO to $\pm$1\% at NLO, and we have a K-factor of
1.36 for $\mu_F=\mu_R=\mu_0$. The main difference with respect to
the ``\wwa'' case is the numerical contribution of the real corrections,
which amounts to 20 \% of the total
contribution or less.

\begin{figure}[tbp]
\includegraphics[scale = 1]{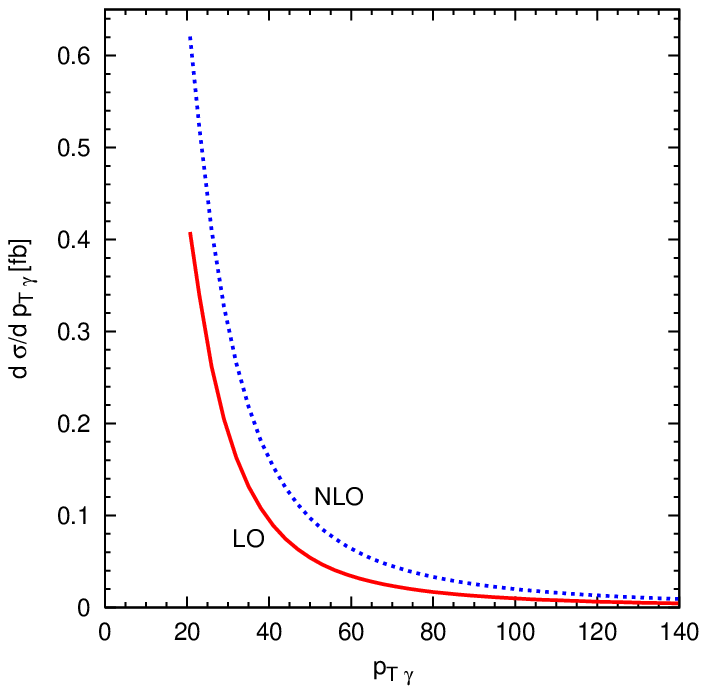}
\includegraphics[scale = 1]{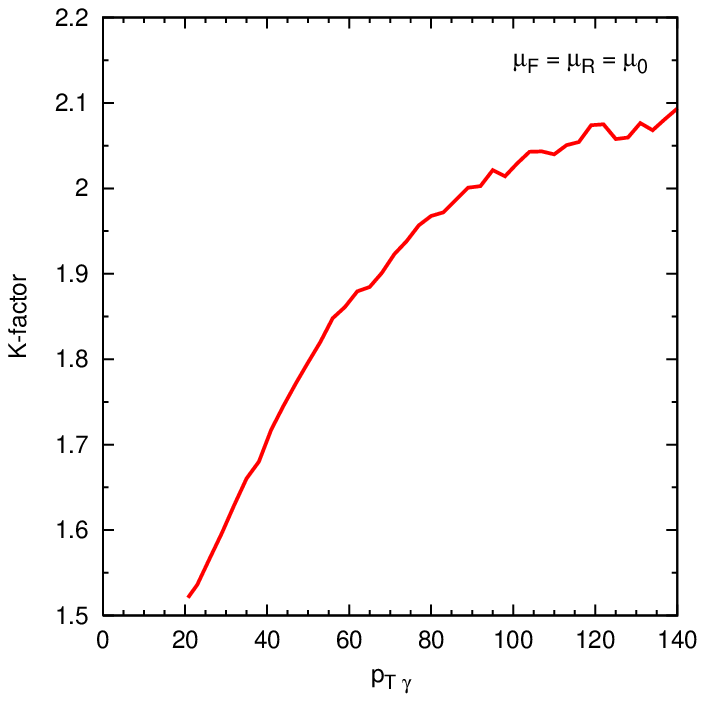}
\caption[]{\label{fig:3}
{\it Left:} {\sl Transverse-momentum distribution of the photon in \wwa\,
  production at the LHC. LO and NLO results are shown for $\mu_F=\mu_R=\mu_0$ 
  and the cuts of Eq.~(\ref{eq:cuts}).} 
{\it Right:} {\sl K-factor for the transverse-momentum distribution of
  the photon as defined in Eq.(\ref{eq:kfactor}).}}
\end{figure}

\begin{figure}[tbp]
\includegraphics[scale = 1]{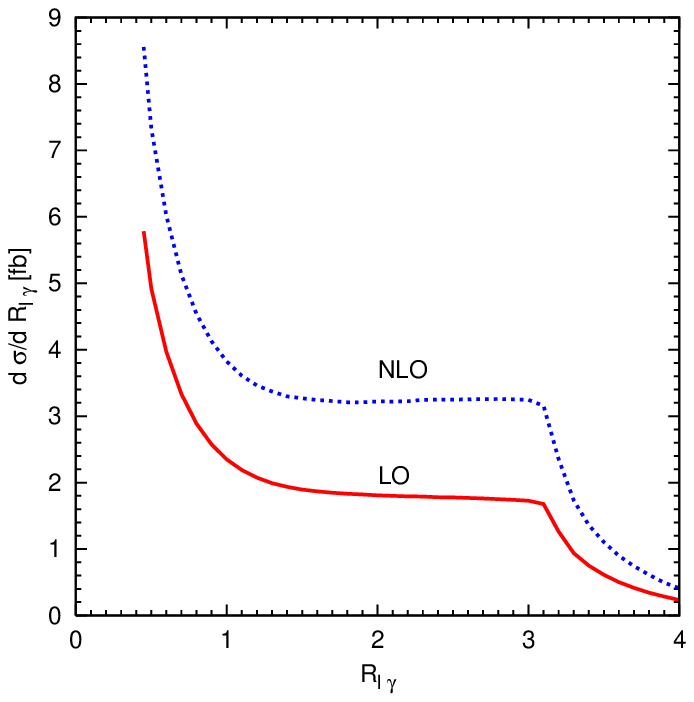}
\includegraphics[scale = 1]{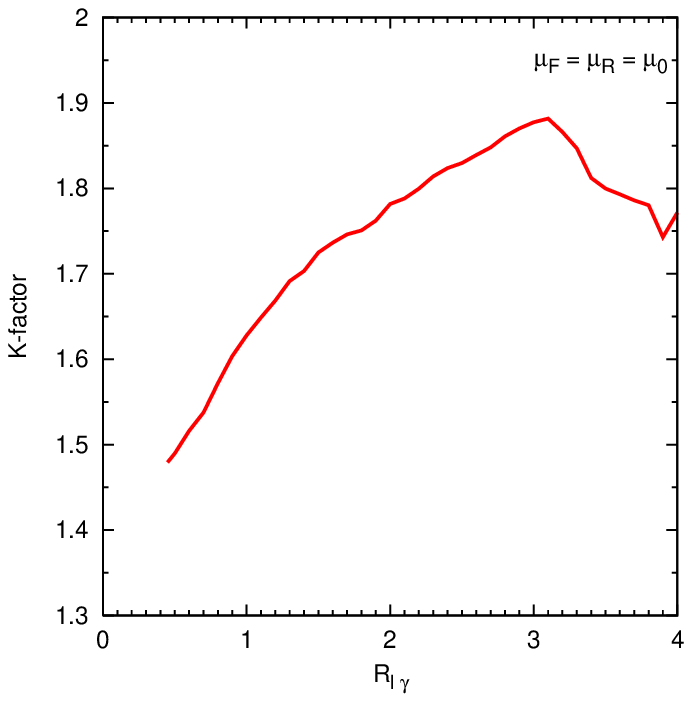}
\caption[]{\label{fig:4}
{\it Left:} {\sl Separation between the photon and the softest lepton
  in \wwa\, production at the LHC, at LO and NLO with 
  $\mu_F=\mu_R=\mu_0$ and the cuts of
  Eq.~(\ref{eq:cuts}).} 
{\it Right:} {\sl K-factor for the separation between the photon and the
softest lepton as defined in Eq.(\ref{eq:kfactor}).}}
\end{figure}

The NLO contributions cannot be described by a simple rescaling of the LO
results since the corrections show a strong dependence on the phase
space region under investigation. As practical examples, we plot two
differential distributions at LO and NLO together with the associated
K-factor, defined as 
\begin{align}\label{eq:kfactor}
K = \frac{d \sigma^{NLO}/ dx}{d \sigma^{LO} /dx} \,\, .
\end{align}
The observables, $x$, considered are the transverse-momentum of the
photon and the separation between the softest (lowest-$p_T$) lepton and
the photon in \wwa\, production at the LHC.
In Figure~\ref{fig:3} we show the transverse-momentum distribution of the
on-shell photon: the corresponding K-factor increases from 1.5 at
$p_{T\gamma}=20$~GeV to 2.1 at $p_{T\gamma}=140$~GeV. 
In Figure \ref{fig:4}, we consider the separation between
the photon and the softest lepton and a similar variation of the
K-factor in the range $1.5-1.9$ is observed.

%
%
\section{Conclusions}
\label{sec:concl}
We have calculated the NLO QCD corrections to the processes $pp,p\bar p\to$
\wwa$+X$ and $pp,p\bar p\to$ \zza$+X$ with full leptonic decays of the 
$W$ and $Z$ bosons. These processes can be relevant both for New Physics 
searches (as
a background) and for the measurement of quartic gauge couplings (as a
signal) at the LHC.

Our numerical results show sizeable cross section increases with respect 
to the LO calculation, ranging from 40\% to as much as 100\% in certain 
regions of phase space. \wwa\, and \zza\, production at the LHC provide 
additional examples of cross sections whose theoretical errors at LO are 
substantially underestimated by considering scale variations only: the 
LO factorization scale variation is much smaller than the NLO correction.
Remaining NLO scale variations are at the 10\% level (e.g. $\pm 6\%$ for
the integrated \wwa\, production cross section at the LHC when varying
$\mu_R=\mu_F=\mu$ by a factor of 2 around the reference scale 
$\mu=Q=m_{WW\gamma}$).

Given the size of the higher-order corrections and, in particular, their
strong dependence on the observable and on different phase space regions
under investigation, a fully-exclusive NLO parton Monte Carlo for \wwa\,
and \zza\, production is required to match the expected precision
of the LHC measurements. We plan to incorporate this and other processes
with a final state photon into the \vbfnlo \, package in the near
future.   
%
\section*{Acknowledgments}
We would like to thank Christoph Englert and Michael Rauch for helpful
discussions, and Malgorzata Worek, for the comparison
with {\tt HELAC}. This research was supported in part by the Deutsche
Forschungsgemeinschaft via the Sonderforschungsbereich/Transregio
SFB/TR-9 ``Computational Particle Physics''.
F.C.~acknowledges a postdoctoral fellowship of the
Generalitat Valenciana (Beca Postdoctoral d'Excel.l\`encia) and partial support by European FEDER and Spanish MICINN under grant FPA2008-02878. The Feynman
diagrams in this paper were drawn using Axodraw~\cite{axo}. 
%
%

\end{document}